# ON SOME NEW NEIGHBOURHOOD DEGREE BASED INDICES

Sourav Mondal[a], Nilanjan De[b*] and Anita Pal[a]

[a]Department of Mathematics, National Institute of Technology, Durgapur, India.

[b]Department of Basic Sciences and Humanities (Mathematics), Calcutta Institute of Engineering and Management, Kolkata, India.

**Abstract:** In this paper, four novel topological indices named as neighbourhood version of forgotten topological index($F_N$), modified neighbourhood version of Forgotten topological index($F_N^*$), neighbourhood version of second Zagreb index ($M_2^*$) and neighbourhood version of hyper Zagreb index ($HM_N$) are introduced. Here the relatively study depends on the structure-property regression analysis is made to test and compute the chemical applicability of these indices for the prediction of physicochemical properties of octane isomers. Also it is shown that these newly presented indices have well degeneracy property in comparison with other degree based topological indices. Some mathematical properties of these indices are also discussed here.

## Introduction

Throughout this article we use only molecular graph[1,2], a connected graph having no loops and parallel edges. In molecular graph nodes and edges correspond to the atoms and chemical bonds of compound, respectively. Let $G$ be a chemical graph containing $V(G)$ and $E(G)$ as vertex set and edge set respectively. The degree of a vertex $v$ on a graph $G$, denoted by $deg_G(v)$, is the total number of edges associated with $v$. Let $N_G(v)$ denotes the set of neighbours of the vertex $v$. In chemical graph theory, topological indices play a leading role specifically in the quantitative structure property relationships and quantitative structure activity relationship modelling[3]. A topological index is a numeric value that is graph invariant. A real valued mapping considering graphs as arguments is called a graph invariant if it gives same value to isomorphic graphs. The order (total count of nodes) and size (total count of edges) of a graph are examples of two graph in variants. In chemical graph theory, the graph invariants are named as topological indices. The idea of topological indices was initiated when the eminent chemist Harold Wiener found the first topological index, known as Wiener index[4] in 1947 for searching boiling points of alkanes. One of the topological indices invented on initial stage is the so called Zagreb index first presented by Gutman and Trinajstić[5,6], where they investigated how the total energy of $\pi$-electron depends on the structure of molecules and it was discussed in details. The first $(M_1(G))$ and the second $(M_2(G))$ Zagreb indices for a molecular graph $G$ are defined as follows:

$$M_1(G) = \sum_{v \in V(G)} deg_G(v)^2 = \sum_{uv \in E(G)} [deg_G(u) + deg_G(v)],$$

$$M_2(G) = \sum_{uv \in E(G)} deg_G(u) deg_G(v).$$

---

* N. De , *e-mail*: de.nilanjan@rediffmail.com

For more discussion on these indices, inquisitive readers are referred the papers[7-16]. Furtula et al.[17] introduced the forgotten topological indices as follows:

$$F(G) = \sum_{v \in V(G)} deg_G(v)^3 = \sum_{uv \in E(G)} [deg_G(u)^2 + deg_G(v)^2].$$

For more discussion on this index readers are referred[18-20]. Following the first Zagreb index present authors[21] introduced a degree based topological index named as the Neighbourhood Zagreb index $(M_N)$ which is defined as follows :

$$\delta_G(v) = \sum_{u \in N_G(v)} deg_G(u),$$

$$M_N(G) = \sum_{v \in V(G)} \delta_G(v)^2.$$

Inspiring from the Zagreb and Forgotten topological indices we present here four new topological indices named as neighbourhood version of forgotten topological index $(F_N)$, modified neighbourhood version of Forgotten topological index $(F_N^*)$, neighbourhood version of second Zagreb index $(M_2^*)$ and neighbourhood version of hyper Zagreb index $(HM_N)$ which are defined as follows:

$$F_N(G) = \sum_{v \in V(G)} \delta_G(v)^3,$$

$$F_N^*(G) = \sum_{uv \in E(G)} [\delta_G(u)^2 + \delta_G(v)^2],$$

$$M_2^*(G) = \sum_{uv \in E(G)} [\delta_G(u)\delta_G(v)],$$

$$HM_N(G) = \sum_{uv \in E(G)} [\delta_G(u) + \delta_G(v)]^2.$$

The objective of this work is to discuss some mathematical properties and check the chemical applicability of these newly introduced indices. Here we find the correlation coefficients of the newly designed indices and some well-established indices with acentric factor and entropy for octane isomers. In addition we investigate the degeneracy of the novel indices.

**Preliminaries**

In this section, we obtain some mathematical properties of the newly introduced indices. Applying some standard lemma, we compute some bounds of the aforesaid novel indices. We start with the following lemma:

**Lemma 1.** For a graph $G$, we have

(i) $\sum_{u \in V(G)}[\delta_G(u)] = M_1(G)$,

(ii) $\sum_{uv \in E(G)}[\delta_G(u) + \delta_G(v)] = 2M_2(G)$.

**Lemma 2.** (Cauchy-Schwartz inequality)[22] Let $x_i$ and $y_i$ be real numbers for all $1 \leq i \leq n$. Then

$$(\sum_{i=1}^{n} x_i y_i)^2 \leq (\sum_{i=1}^{n} x_i^2)(\sum_{i=1}^{n} y_i^2). \tag{1}$$

Equality holds $iff$ $x_i = k y_i$ for some constant $k$ and for each $1 \leq i \leq n$.

**Proposition 1.** Let $G$ be a graph with $m$ edges, whose second Zagreb index is $M_2(G)$, then we have

$$HM_N(G) \geq \frac{4M_2^2(G)}{m},$$

with equality holds $iff$ $\delta_G(u) + \delta_G(v) = k$ for some constant $k$, $\forall uv \in E(G)$.

**Proof.** In (1), considering $x_i = \delta_G(u) + \delta_G(v), y_i = 1$, we have

$$\left(\sum_{uv \in E(G)} [\delta_G(u) + \delta_G(v)]\right)^2 \leq \sum_{uv \in E(G)} [\delta_G(u) + \delta_G(v)]^2 \sum_{uv \in E(G)} 1^2,$$

applying the definition of $HM_N$ and lemma 1, we get the required result. From lemma 2, it is clear that equality holds $iff$ $\delta_G(u) + \delta_G(v) = k$ for some constant $k$, $\forall uv \in E(G)$.

**Lemma 3.**[23] Let $(x_1, x_2, \ldots, x_n)$ be positive $n$-tuple such that there exists positive number $A, a$ satisfying $0 \leq a \leq x_i \leq A$, then we have

$$\frac{n \sum_{i=1}^{n} x_i^2}{\left(\sum_{i=1}^{n} x_i\right)^2} \leq \frac{1}{4}\left(\sqrt{\frac{A}{a}} + \sqrt{\frac{a}{A}}\right)^2, \tag{2}$$

where equality holds $iff\ a = A$ or $q = \frac{\frac{A}{a}}{\frac{A}{a}+1}n$ is an integer and $q$ of the numbers $x_i$ coincide with $a$ and the remaining $(n-q)$ of the $x_i'$ s coincide with $A(\neq a)$.

For a graph $G$ consider
$$\Delta_N = \max\{\delta_G(v): u \in V(G)\},$$
$$\delta_N = \min\{\delta_G(v): u \in V(G)\}.$$

Now putting $a = 2\delta_N$, $A = 2\Delta_N$, $x_i = \delta_G(u) + \Delta_G(v)$ in (2) and using lemma 1, we have the following proposition.

**Proposition 2.** Let $G$ be a graph with $m$ edges, whose second Zagreb index is $M_2(G)$, then we have

$$HM_N(G) \leq \frac{M_2(G)^2}{m} \frac{(\Delta_N + \delta_N)^2}{\Delta_N \delta_N},$$

where equality holds $iff\ \Delta_N = \delta_N$ or $q = \frac{\frac{\Delta_N}{\delta_N}}{\frac{\Delta_N}{\delta_N}+1}m$, is an integer and $q$ of the numbers $x_i$ coincide with $\delta_N$ and the remaining $(m-q)$ of the $x_i'$ s coincide with $\Delta_N(\neq \delta_N)$.

**Lemma 4.** Let $\vec{x} = (x_1, x_2, ..., x_n)$ and $\vec{y} = (y_1, y_2, ..., y_n)$ be sequence of real numbers. Also let $\vec{z} = (z_1, z_2, ..., z_n)$ and $\vec{w} = (w_1, w_2, ..., w_n)$ be non-negative sequences. Then,

$$\sum_{i=1}^{n} w_i \sum_{i=1}^{n} z_i x_i^2 + \sum_{i=1}^{n} z_i \sum_{i=1}^{n} w_i y_i^2 \geq 2 \sum_{i=1}^{n} z_i x_i \sum_{i=1}^{n} w_i y_i. \qquad (3)$$

In particular, if $z_i$ and $w_i$ are positive, then the equality holds $iff\ \vec{x} = \vec{y} = \vec{k}$, where $\vec{k} = (k, k, ..., k)$, a constant sequence.

**Proposition 3.** For any graph $G$ with Neighbourhood Zagreb index and first Zagreb index $M_N(G)$ and $M_1(G)$ respectively, we have

$$F_N(G) \geq 2M_N(G) - M_1(G), \qquad (4)$$

where equality holds $iff\ G$ is $P_2$.

**Proof.** Considering $x_i = \delta_G(u)$, $y_i = 1$, $z_i = \delta_G(u)$, $w_i = 1$ in (3), we obtain

$$\sum_{i=1}^{n} 1 \sum_{u \in V(G)} \delta_G(u)^3 + \sum_{u \in V(G)} \delta_G(u) \sum_{i=1}^{n} 1 \geq 2 \sum_{u \in V(G)} \delta_G(u)^2 \sum_{i=1}^{n} 1. \quad (5)$$

After using the definition of $F_N$ and $M_N$ indices and applying lemma 1, we obtain the required result. According to the lemma 3, the equality in (4) holds $iff\ \delta_G(u) = 1\ \forall\ u \in V(G), i.e.\ G$ is $P_2$. Hence the proof.

**Lemma 5.** (Radon's inequality) If $a_i, b_i > 0, i = 1, 2, \ldots, n, p > 0$, then

$$\frac{\sum_{i=1}^{n} a_i^{p+1}}{\sum_{i=1}^{n} b_i^{p}} \geq \frac{\left(\sum_{i=1}^{n} a_i\right)^{p+1}}{\left(\sum_{i=1}^{n} b_i\right)^{p}}, \quad (6)$$

where equality holds $iff\ a_i = k b_i$ for some constant $k, \forall i = 1, 2, \ldots, n$.

For a graph $G$, considering $a_i = \delta_G(u), b_i = 1, p = 2$, in (6), we have the following proposition.

**Proposition 4.** For any graph with n vertices, we have

$$F_N(G) \geq \frac{M_1(G)^3}{n}, \quad (7)$$

where equality holds $iff\ G$ is regular or complete bipartite graph.

**Proposition 5.** Let $G$ be a graph, whose first and neighbourhood Zagreb indices are $M_1(G)$ and $M_N(G)$ respectively. Then

$$F_N(G) \geq \frac{M_N(G)^2}{M_1(G)}. \quad (8)$$

**Proof.** Let $G$ be a graph and $u \in V(G)$. The weighted averages of $\delta_G(u)$ and squares of $\delta_G(u)$ are

$$\langle d \rangle_w = \frac{\sum_{u \in V(G)} w(u) \delta_G(u)}{\sum_{u \in V(G)} w(u)},$$

$$\langle d^2 \rangle_w = \frac{\sum_{u \in V(G)} w(u) \delta_G(u)^2}{\sum_{u \in V(G)} w(u)},$$

where $w(u)$ is weight corresponding to the vertex $u$ of $G$. For any non-negative weight, $\langle d^2 \rangle_w \geq (\langle d \rangle_w)^2$. Choosing $w(u) = \delta_G(u)$ and using definitions of $F_N(G), M_N(G)$ and $M_1(G)$, we obtain the bound (8).

**Chemical significance of the newly introduced indices**

According to the report of the IAMC (International Academy of Mathematical Chemistry), the chemical applicability of a topological index can be evaluated by regression analysis. Naturally 18 octane isomers are helpful for such investigation, since the number of the structural isomers of octane is large (18) enough to create the statistical perfection faithful. Furtula et al.[17] shown that $M_1$ and $F$ yield correlation coefficient greater than 0.95 with acentric factor and entropy for octane isomers. Also a simple linear model $(M_1 + \lambda F)$, where $\lambda$ is varied from -20 to 20 is designed to improve the predictive ability of these indices. De et al. computed that the correlation coefficient of F-coindex for octane isomers in case of the logarithm of the octanol-water partition coefficient $(P)$ is 0.966. In a recent work[24] the application possibilities of various graph irregularity indices for the prediction of physicochemical properties are described. We find the correlation of different physiochemical properties with $F_N, F_N{}^*, M_2{}^*$ and $HM_N$ of octane isomers and good results are obtained in case acentric factor (Acent Fac.) and entropy (S) which are shown in this report (Table 2). The correlations of acentric factor and entropy with some well-known degree based topological indices are also investigated in Table 3. The results are also shown graphically in Figure 1 and Figure 2. The datas of octane isomers (Table1) are collected from www.moleculardescriptors.eu/dataset/dataset.htm. Thus the newly introduced indices can help to predict the entropy and acentric factor with powerful accuracy.

**Table 1.** Experimental values of the acentric factor, entropy $(S)$ and the corresponding values of $F_N, F_N{}^*, M_2{}^*$, and $HM_N$.

| Molecule name | Acent Fac. | S | $F_N(G)$ | $F_N{}^*(G)$ | $M_2{}^*(G)$ | $HM_N(G)$ |
|---|---|---|---|---|---|---|
| n-octane | 0.397898 | 111.67 | 326 | 172 | 84 | 340 |
| 2-methyl heptane | 0.377916 | 109.84 | 406 | 202 | 98 | 398 |
| 3-methyl heptane | 0.371002 | 111.26 | 448 | 224 | 106 | 436 |
| 4-methyl heptane | 0.371504 | 109.32 | 472 | 228 | 107 | 442 |
| 3-ethyl hexane | 0.362472 | 109.43 | 520 | 252 | 115 | 482 |
| 2,2-dimethyl hexane | 0.339426 | 103.42 | 632 | 282 | 132 | 538 |
| 2,3-dimethyl hexane | 0.348247 | 108.02 | 582 | 282 | 129 | 540 |
| 2,4-dimethyl hexane | 0.344223 | 106.98 | 558 | 258 | 121 | 500 |
| 2,5-dimethyl hexane | 0.35683 | 105.72 | 486 | 232 | 113 | 458 |
| 3,3-dimethyl hexane | 0.322596 | 104.74 | 728 | 324 | 148 | 620 |
| 3,4-dimethyl hexane | 0.340345 | 106.59 | 630 | 306 | 136 | 578 |

**Table 1. Continued**

| | | | | | | |
|---|---|---|---|---|---|---|
| 2-methyl-3-ethyl pentane | 0.332433 | 106.06 | 666 | 312 | 137 | 586 |
| 3-methyl-3-ethyl pentane | 0.306899 | 101.48 | 806 | 374 | 163 | 700 |
| 2,2,3-trimethyl pentane | 0.300816 | 101.31 | 850 | 384 | 171 | 726 |
| 2,2,4-trimethyl pentane | 0.30537 | 104.09 | 778 | 312 | 147 | 606 |
| 2,3,3-trimethyl pentane | 0.293177 | 102.06 | 874 | 408 | 179 | 766 |
| 2,3,4-trimethyl pentane | 0.317422 | 102.39 | 728 | 342 | 151 | 644 |
| 2,2,3,3-tetramethyl butane | 0.255294 | 93.06 | 1070 | 488 | 217 | 922 |

**Table 2**. Correlation coefficient of $F_N, F_N^*, M_2^*$ and $HM_N$ with acentric factor and entropy $(S)$.

| | $F_N(G)$ | $F_N^*(G)$ | $M_2^*(G)$ | $HM_N(G)$ |
|---|---|---|---|---|
| Acent Fac. | -0.99457 | -0.97547 | -0.98533 | -0.98049 |
| S | -0.93831 | -0.93164 | -0.94809 | -0.93784 |

**Table 3**. Correlation coefficient of $M_1, M_2$, and $F$ with acentric factor and entropy $(S)$.

| | $M_1(G)$ | $M_2(G)$ | $F(G)$ |
|---|---|---|---|
| Acent Fac. | -0.97306 | -0.98642 | -0.96505 |
| S | -0.95429 | -0.94169 | -0.95272 |

Now, we depict the correlations discussed above in the following figures.

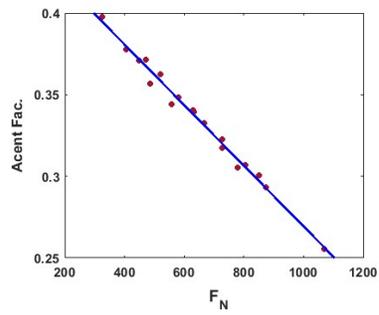
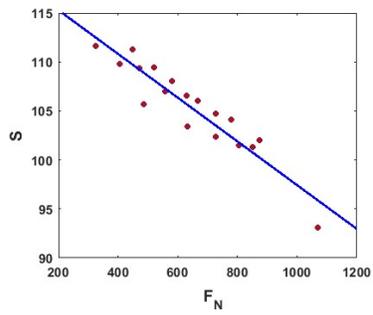
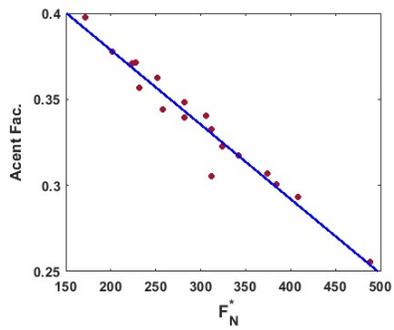
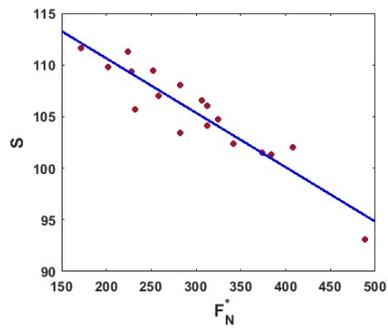
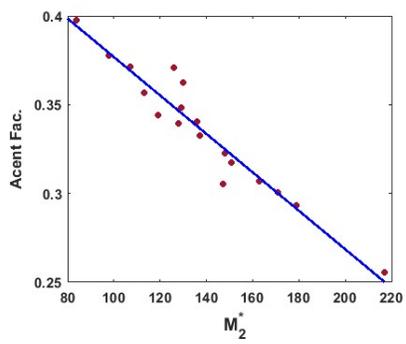
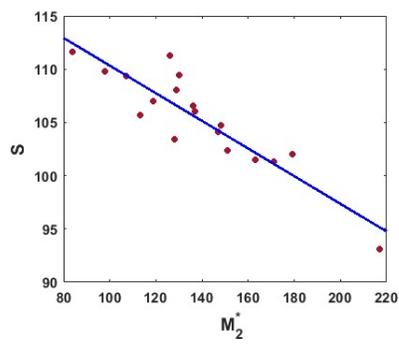
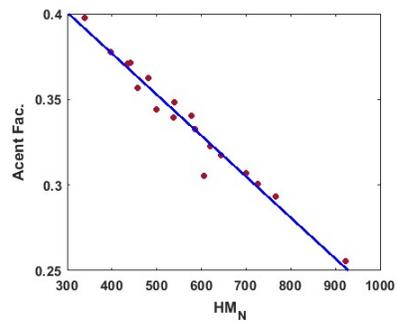
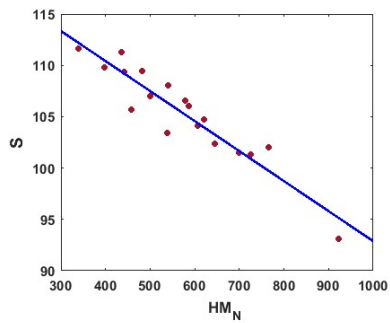

**Figure 1**. Correlations of acentric factor and entropy ($S$) with the newly introduced indices for octane isomers.

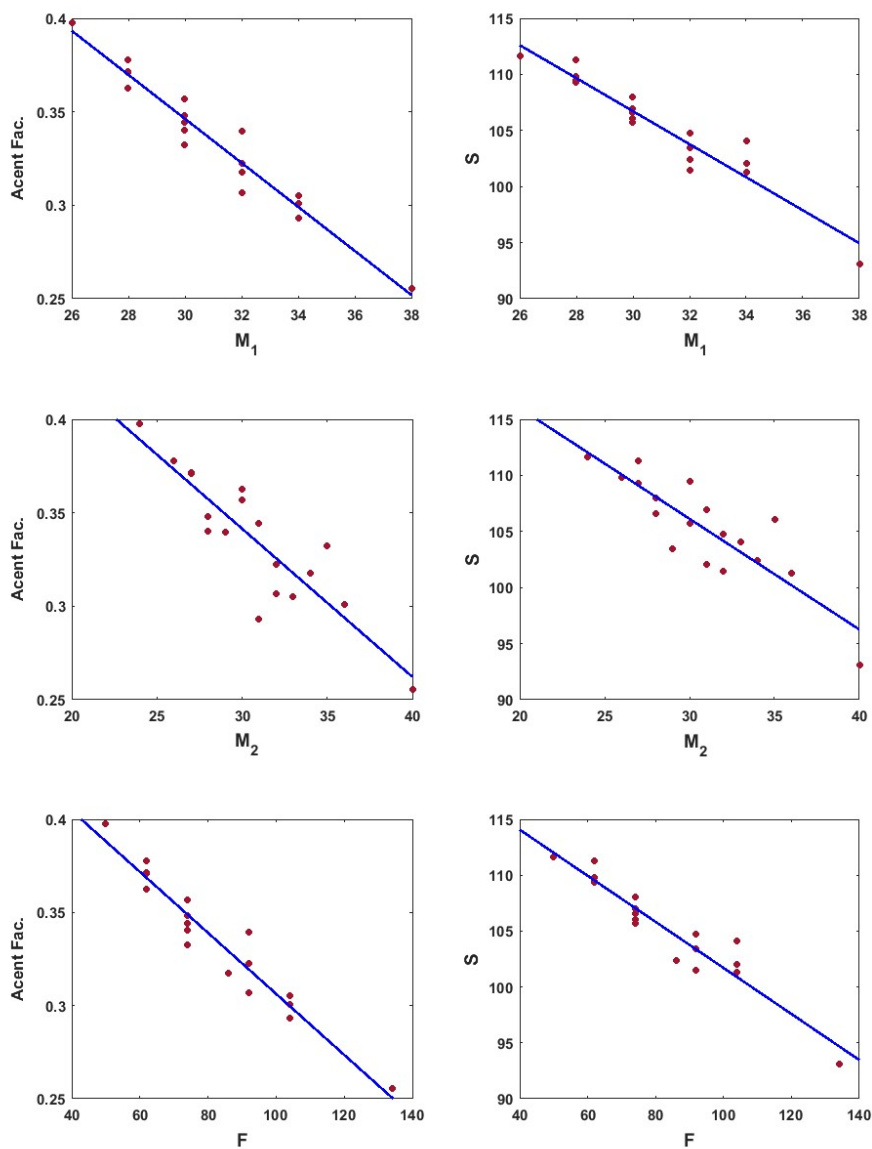

**Figure 2.** Correlations of acentric factor and entropy ($S$) with some well established indices ($M_1, M_2$, and $F$) for octane isomers.

The aim of molecular descriptors is to encode the structural characteristics of a molecule to the greatest extent possible. Ideally, a molecular descriptor should distinguish between two different structural formulae. A major drawback of most topological indices is their degeneracy, i.e., two or more isomers possess the same topological index. Topological indices having high discriminating power captures more structural information. We use the measure of degeneracy known as sensitivity introduced by Konstantinova[25], which is defined as follows:

$$S_I = \frac{N - N_I}{N},$$

where $N$ is the total number of isomers considered and $N_I$ is the number of them that cannot be distinguished by the topological index $I$. As $S_I$ increases, the isomer-discrimination power of topological indices increases. The vertex degree based topological indices have more discriminating power in comparison with other classes of molecular descriptors. For octane isomers, the newly introduced indices exhibit good response among other investigated degree based indices (Table 4).

**Table 4.** Measure of sensitivity ($S_I$) of different indices for octane isomers.

| Indices | Sensitivity ($S_I$) |
|---|---|
| $M_2^*$ | **1.000** |
| $HM_N$ | **1.000** |
| $F_N$ | **0.944** |
| $F_N^*$ | **0.889** |
| Connectivity index ($\chi$) | 0.889 |
| Hyper Zagreb index ($HM$) | 0.833 |
| Second Zagreb index ($M_2$) | 0.722 |
| Hosoya index ($Z$) | 0.778 |
| Forgotten topological index ($F$) | 0.389 |
| First Zagreb index ($M_1$) | 0.333 |

Also these new indices have very good correlations with some well-established degree based topological indices (Table 5) which can predict physiochemical properties with high accuracy. Thus we can say that these novel indices are chemically significant.

**Table 5.** Correlation coefficients of $F_N, F_N^*, M_2^*$ and $HM_N$ with some other indices.

|        | $F_N$ | $F_N^*$, | $M_2^*$ | $HM_N$ | $M_1$ | $M_2$ | $F$ |
|--------|-------|----------|---------|--------|-------|-------|-----|
| $F_N$  | 1     |          |         |        |       |       |     |
| $F_N^*$, | 0.987 | 1      |         |        |       |       |     |
| $M_2^*$ | 0.992 | 0.996   | 1       |        |       |       |     |
| $HM_N$ | 0.989 | 0.999   | 0.999   | 1      |       |       |     |
| $M_1$  | 0.961 | 0.925   | 0.952   | 0.936  | 1     |       |     |
| $M_2$  | 0.991 | 0.997   | 0.998   | 0.998  | 0.948 | 1     |     |
| $F$    | 0.957 | 0.919   | 0.948   | 0.931  | 0.996 | 0.939 | 1   |

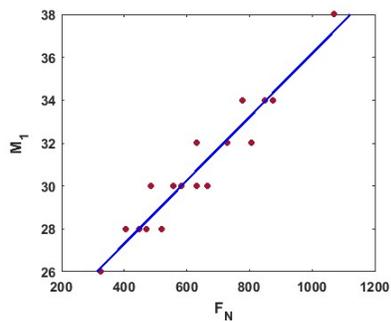 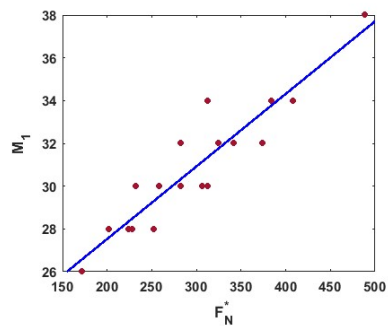

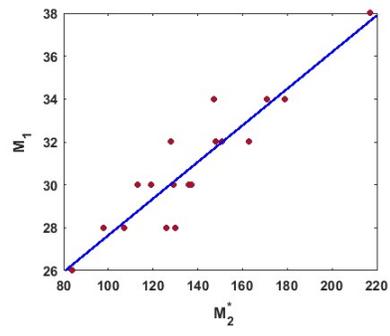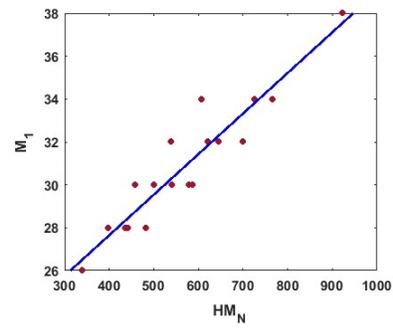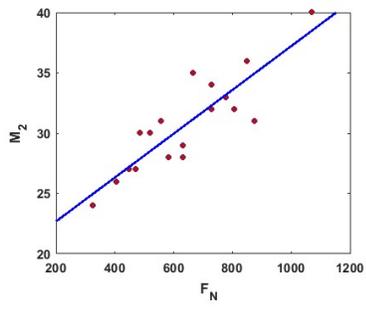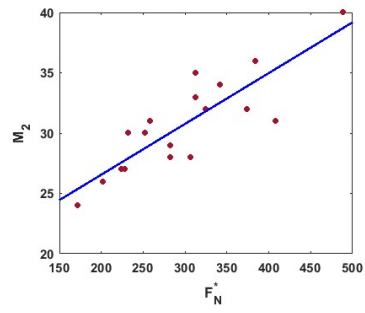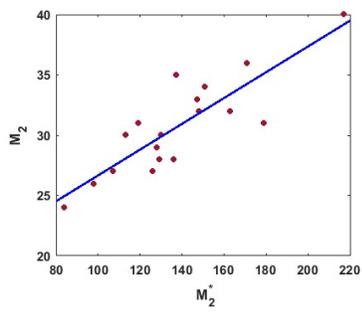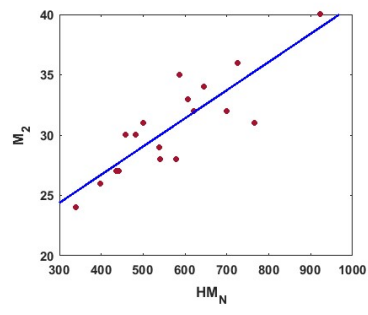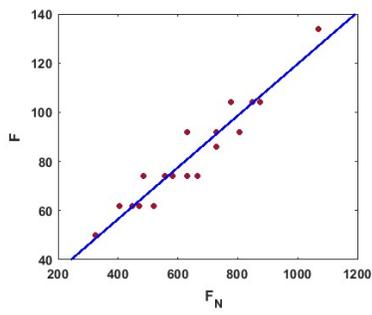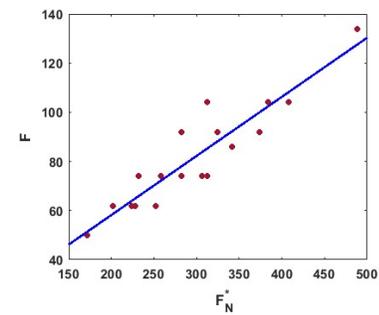

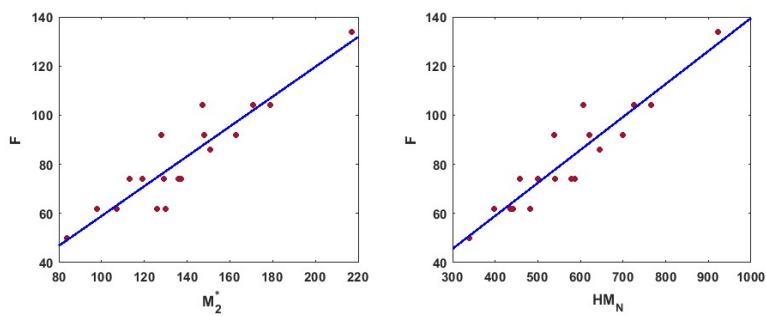

**Figure 3.** Correlations of some well-established degree based indices ($M_1, M_2$, and $F$) with novel indices.

### Conclusion

In this report, we have introduced some new topological indices. Some mathematical properties of the newly proposed topological indices are discussed. Their chemical applicability is also investigated here. These indices have significant correlation with acentric factor and entropy in comparison with $M_1, M_2,$ and $F$, shown in Table 2 and Table 3. Also Table 4 exhibits their supremacy in discriminative power in comparison to the other well-known investigated indices. Thus the four novel indices $F_N, F_N^*, M_2^*$ and $HM_N$ deserve to be considered as applicable topological indices. We have correlated indices among themselves and with some other well-known degree-based topological indices in Table 5. From the correlations among the novel indices, it is clear that $F_N$ and $F_N^*$ have good quality among four indices. For further research, these indices can be computed for various graph operations and some composite graphs and networks.

### Acknowledgement

The first author is very obliged to the Department of Science and Technology (DST), Government of India for the Inspire Fellowship [IF170148].